\documentclass[sigconf,screen]{acmart}

\PassOptionsToPackage{warn}{textcomp}

\usepackage{xcolor}[table]
\usepackage{soul}
\usepackage[utf8]{inputenc}
\usepackage{comment}
\usepackage{color}
\usepackage{framed}
\usepackage{epsfig}
\usepackage{graphicx}
\usepackage{amsmath}
\usepackage{amssymb}

\usepackage{url}
\usepackage{multirow,makecell}
\usepackage{enumitem}
\usepackage{moresize}
\usepackage{epstopdf}
\usepackage{array}

\usepackage[T1]{fontenc}

\usepackage{microtype}
\usepackage{balance}

\usepackage{tcolorbox}

\usepackage{colortbl}
\definecolor{light-gray}{gray}{0.92}
\usepackage{multirow}

\usepackage{wrapfig}
\usepackage{subfigure}
\usepackage[hang,flushmargin]{footmisc}

\usepackage[belowskip=-10pt,aboveskip=5pt]{caption}

\usepackage{xspace}
\newcommand{\wrt}{\hbox{\emph{w.r.t.}}\xspace}

\let\OLDthebibliography\thebibliography
\renewcommand\thebibliography[1]{
  \OLDthebibliography{#1}

  \setlength{\parskip}{3.0pt}

  \setlength{\itemsep}{0pt plus 1ex}
}

\newcommand{\etal}{\textit{et al}.\xspace}
\newcommand{\ie}{\textit{i}.\textit{e}.\xspace}
\newcommand{\eg}{\textit{e}.\textit{g}.\xspace}

\title{\emph{DeepGauge}: Multi-Granularity Testing Criteria for Deep Learning Systems}

\author{
Lei Ma$^{1,3\ast}$, 
Felix Juefei-Xu$^2$, 
Fuyuan Zhang$^3$,
Jiyuan Sun$^4$, 
Minhui Xue$^3$,
Bo Li$^5$\\
Chunyang Chen$^6$,
Ting Su$^3$,
Li Li$^6$,
Yang Liu$^3$,
Jianjun Zhao$^4$,
and Yadong Wang$^1$}
\thanks{{$^{\ast}$} Lei Ma is the corresponding author. Email: malei@hit.edu.cn.}
\affiliation{$^1$Harbin Institute of Technology, China $^2$Carnegie Mellon University, USA $^3$Nanyang Technological University, Singapore\\ $^4$Kyushu University, Japan
 $^5$University of Illinois at Urbana–Champaign, USA $^6$Monash University, Australia}

\setcopyright{acmcopyright}
\acmPrice{15.00}
\acmDOI{10.1145/3238147.3238202}
\acmYear{2018}
\copyrightyear{2018}
\acmISBN{978-1-4503-5937-5/18/09}
\acmConference[ASE '18]{Proceedings of the 2018 33rd ACM/IEEE International Conference on Automated Software Engineering}{September 3--7, 2018}{Montpellier, France}
\acmBooktitle{Proceedings of the 2018 33rd ACM/IEEE International Conference on Automated Software Engineering (ASE '18), September 3--7, 2018, Montpellier, France}


\begin{document}

\renewcommand{\shortauthors}{Ma, Xu, Zhang, Sun, Xue, Li, Chen, Su, Li, Liu, Zhao, and Wang}

\begin{abstract}
Deep learning (DL) defines a new data-driven programming paradigm that constructs the internal system logic of a crafted neuron network through a set of training data. We have seen wide adoption of DL in many safety-critical scenarios. However, a plethora of studies have shown that the state-of-the-art DL systems suffer from various vulnerabilities which can lead to severe consequences when applied to real-world applications. Currently, the testing adequacy of a DL system is usually measured by the accuracy of test data. 
Considering the limitation of accessible high quality test data, good accuracy performance on test data can hardly provide confidence to the testing adequacy and generality of DL systems. 
Unlike traditional software systems that have clear and controllable logic and functionality, the lack of interpretability in a DL system makes system analysis and defect detection difficult, which could potentially hinder its real-world deployment. In this paper, we propose \emph{DeepGauge}, a set of multi-granularity testing criteria for DL systems, which aims at rendering a multi-faceted portrayal of the testbed. The in-depth evaluation of our proposed testing criteria is demonstrated on two well-known datasets, five DL systems, and with four state-of-the-art adversarial attack techniques against DL. The potential usefulness of \emph{DeepGauge} sheds light on the construction of more generic and robust DL systems.
\end{abstract}


\maketitle

\section{Introduction}

Deep learning (DL) systems have gained great popularity in various applications, \eg, speech processing \cite{hinton2012deep}, medical diagnostics~\cite{ciresan2012deep}, image processing \cite{ciregan2012multi}, and robotics \cite{zhang2015towards}. A deep neural network (DNN), as a type of deep learning systems, is the key driving force behind recent success.
However, DNN-based software systems, such as autonomous driving, often exhibit erroneous behaviors that lead to fatal consequences. For example, several accidents \cite{google_crash} have been reported due to autonomous vehicle's failure to handle unexpected/corner-case driving conditions.

One of the trending research areas is to investigate the cause of vulnerability in DL systems by means of generating adversarial test examples for image- and video-based DL systems. Such carefully learned pixel-level perturbations, imperceptible to human eyes, can cause the DL-based classification system to output completely wrong decisions with high confidence \cite{goodfellow2014explaining}. Ever since the inception of adversarial attacks on the DL systems, more and more research has been dedicated to building up strong attackers~ \cite{iclr18-ae-boundary-analysis,iclr18-a-boundary-attack,iclr18-ae-natural,iclr18-ae-spatial}. As a consequence, better defense mechanisms in DL systems against adversarial attacks are in dire need. Various techniques to nullify adversarial attacks and to train a more robust DL system are emerging in recent studies \cite{iclr18-ab-binarized,iclr18-ab-ensemble,iclr18-b-certified,iclr18-b-counter,iclr18-b-defense-gan,iclr18-b-pixeldefend,iclr18-b-random}. Together, research in both realms forms a virtuous circle and blazes a trail for better understanding of how to build more generic and robust DL systems.

However, what is still lacking is a systematic way of gauging the testing adequacy of given DL systems. Current studies focus only on pursuing high accuracy of DL systems as a testing criterion, for which we show several caveats as follows. 
\emph{\textbf{First}}, measuring the software quality from DL output alone is superficial in the sense that fundamental understanding of the DL internal neuron activities and network behaviors is not touched upon. We agree that it could be an indicator of DL system quality and generality, but it is far from complete, and oftentimes unreliable. 
\emph{\textbf{Second}}, a criterion solely based on DL output will rely heavily on how representative the test data are. Having achieved high-performance DL output does not necessarily mean that the system is utmost generic, and achieving low-performance does not indicate the opposite either. A DL model can be immune to many known types of adversarial attacks, but may fail from unseen attacks. This is because such a criterion based only on DL outputs is far from being comprehensive, and it leaves high risks for currently cocooned DL systems to be deployed in the real-world environment where newly evolved adversarial attacks are inescapable. 
\emph{\textbf{Third}}, any DL system that passes systematic testing should be able to withstand all types of adversarial attacks to some extent. Such generality upon various attacks is of vital importance for DL systems to be deployed. But apparently this is not the case, unless we stick to a set of more comprehensive gauging criteria. We understand that even the most comprehensive gauging criteria would not be able to entirely eliminate risks from adversarial attacks. Nevertheless, by enforcing a suitable set of testing criteria, we hope that a DL system could be better tested to facilitate the construction of a more generic and robust deep learning system.

Towards addressing the aforementioned limitations, \emph{a set of} testing criteria is needed, as opposed to the sole criterion based on DL decision output. In addition to being scalable, the proposed criteria will have to monitor and gauge the neuron activities and intrinsic network connectivity at various granularity levels, so that a multi-faceted in-depth portrayal of the DL system and testing quality measures become desirable.

In this work, we are probing this problem from a software engineering and software testing perspective. At a high level, erroneous behaviors appeared in DNNs are analogous to logic bugs in traditional software. However, these two types of software are fundamentally different in their designs. Traditional software represents its logic as control flows crafted by human knowledge, while a DNN characterizes its behaviors by the weights of neuron edges and the nonlinear activation functions (determined by the training data). Therefore, detecting erroneous behaviors in DNNs is different from detecting those in traditional software in nature, which necessitates novel test generation approaches. 

To achieve this goal, the very first step is to precisely define a set of suitable coverage criteria, which can guide test design and evaluate test quality. Despite a number of criteria existing for traditional software, \eg, statement, branch,  data-flow coverage, they completely lose effect in testing DNNs. To the best of our knowledge, the design of testing coverage criteria for DNNs is still at the early stage~\cite{pei2017deepxplore,2018arXiv180304792S}. Without a comprehensive set of criteria, (1) designing tests to cover different learned logics and rules of DNNs is difficult to achieve. Consequently, erroneous behaviors may be missed; (2) evaluating test quality is biased, and the confidence of obtained testing results may be overestimated.
In this paper, we propose \emph{DeepGauge}---a set of testing criteria based on multi-level and -granularity coverage for testing DNNs and measure the testing quality. Our contributions are summarized as follows: 

\begin{itemize}[wide=1pt,leftmargin=10pt]

\item  Our proposed criteria facilitate the understanding of DNNs as well as the test data quality from different levels and angles. In general, we find defects could potentially distribute on both major function regions as well as the corner-case regions of DNNs. Given a set of inputs, our criteria could measure to what extent it covers the main functionality and the corner cases of the neurons, where DL defects could incur. Our evaluation results reveal that the existing test data of a given DL in general skew more towards testing the major function region, with relatively few cases covering the corner-case region.

\item  In line with existing test data of DNNs, we evaluate the usefulness of our coverage criteria as indicators to quantify defect detection ability of test data on DNNs, through generating new adversarial test data using 4 well-known adversarial data generation algorithms (\ie, Fast Gradient Sign Method (FGSM) \cite{goodfellow2014explaining}, Basic Iterative Method (BIM)~\cite{kurakin2017adversarial}, Jacobian-based Saliency Map Attack (JSMA) ~\cite{papernot2016limitations} and Carlini/Wagner attack (CW)~\cite{cw2017}).
The extensive evaluation shows that our criteria can effectively capture the difference between the original test data and adversarial examples, where DNNs could and could not correctly recognize, respectively, demonstrating that a higher coverage of our criteria potentially indicates a higher chance to detect the DNN's defects.

\item The various criteria proposed behave differently on DNNs \wrt network complexity and dataset under analysis. Altogether, these criteria can potentially help us gain insights of testing DNNs. By providing these insights, we hope that both software engineering and machine learning communities can benefit from applying new criteria for gauging the testing quality of the DNNs to gain confidence towards constructing generic and robust DL systems.

\end{itemize}

To the best of our knowledge, this is among the earliest studies to propose multi-granularity testing criteria for DL systems, which are mirrored by the test coverage in traditional software testing.

\section{Related Work}
In this section, we attempt to review the most relevant work in three aspects: testing, verification, and security of DL systems.

\subsection{Testing of DL Systems}
Traditional practices in measuring machine learning systems mainly rely on probing their accuracy on test inputs which are randomly drawn from manually labeled datasets and \emph{ad hoc} simulations \cite{witten2016data}. However, such black-box testing methodology may not be able to find various kinds of corner-case behaviors that may induce unexpected errors~\cite{goodfellow2017challenge}.
Wicker \etal~\cite{DBLP:journals/corr/abs-1710-07859} recently proposed a Scale Invariant Feature Transform feature guided black-box testing and showed its competitiveness with CW and JSMA along this direction.

Pei \etal~\cite{pei2017deepxplore} proposed a white-box differential testing algorithm for systematically finding inputs that can trigger inconsistencies between multiple DNNs. They introduced neuron coverage for measuring how much of the internal logic of a DNN has been tested.% which we think is a very important work on testing DLs.
However, it still exhibits several caveats as discussed in Section~\ref{sec:deepxplore}.
DeepTest \cite{DBLP:journals/corr/abs-1708-08559} investigates a basic set of image transformations~(\eg, scaling, shearing, and rotation) from OpenCV and shows that they are useful to detect defects in DNN-driven autonomous cars. Along this direction, DeepRoad~\cite{2018arXiv180202295Z} uses input image scene transformation and shows its potentiality with two scenes~(\ie, snowy and rainy) for autonomous driving testing. The scene transformation is obtained through training a generative adversarial network~(GAN) with a pair of collected training data that cover the statistical features of the two target scenes.

Compared with traditional software, the dimension and potential testing space of a DNN is often quite large. DeepCT~\cite{ma2018combinatorial} adapts the concept of combinatorial testing, and proposes a set of coverage based on the neuron input interaction for each layer of DNNs, to guide test generation towards achieving reasonable defect detection ability with a relatively small number of tests. 
Inspired by the MC/DC test criteria in traditional software~\cite{KellyJ.:2001:PTM:886632}, Sun \etal~\cite{2018arXiv180304792S} proposed a set of adapted MC/DC test criteria for DNNs, and show that generating tests guided by the proposed criteria on small scale neural networks~(consisting of \texttt{Dense} layers with no more than $5$ hidden layers and $400$ neurons) exhibits higher defect detection ability than random testing. However, whether MC/DC criteria scale to real-world-sized DL systems still needs further investigation.
Instead of observing the runtime internal behaviors of DNNs, DeepMutation~\cite{2018arXiv180505206M} proposes to mutate DNNs~(\ie, injecting faults either from the source level or model level) to evaluate the test data quality, which could potentially be useful for test data prioritization in respect of robustness on a given DNN.

Our work of proposing multi-granularity testing coverage for DL systems is mostly orthogonal to the existing work. Compared with the extensive study on traditional software testing, testing DL is still at an early stage. Most existing work on DL testing lacks some suitable criteria to understand and guide the test generation process. 
Since test generation guided by coverage criteria~(\eg, statement coverage, branch coverage)
towards the exploration of diverse software states for defect detection has become the \emph{de facto} standard in traditional software testing~\cite{feldt2008searching,baudry2015multiple,Fraser:2013:WTS:2478542.2478706, Ma:2015:GPR:2916135.2916251}, the study to design suitable testing criteria for DL is desperately demanding. This paper makes an early attempt towards this direction by proposing a set of testing criteria. Our criteria not only can differentiate state-of-the-art adversarial test generation techniques, but also potentially be useful for the measurement of test suite diversity by analyzing the DNNs' internal states from multiple portrayals.
We believe that our proposed criteria set up an important cornerstone and bring a new opportunity to design more effective automated testing techniques guided by testing criteria for DL systems.

\subsection{Verification of DL Systems}

Formal methods can provide formal guarantees about safety and robustness of verified DL systems \cite{DBLP:journals/corr/abs-1710-07859,LA10,GCDKM17,DGCC17,TMDPSM18,KYJS17}. The main concern of formal methods are their scalability for real-world-sized (\eg, $100,000$ neurons or even more) DL systems.

The early work in \cite{LA10} provided an abstraction-refinement approach to checking safety properties of multi-layer perceptrons. Their approach has been applied to verify a network with only $6$ neurons. 
DLV \cite{DBLP:journals/corr/abs-1710-07859} can verify local robustness of DL systems w.r.t. a set of user specified manipulations. 
Reluplex \cite{GCDKM17} is a sound and complete SMT-based approach to verifying safety and robustness of DL systems with \texttt{ReLU} activation functions. The networks verified by Reluplex in \cite{GCDKM17} have $8$ layers and $300$ \texttt{ReLU} nodes. 
DeepSafe \cite{DGCC17} uses Reluplex as its verification engine and has the same scalability problem as Reluplex. 
AI$^{2}$ \cite{TMDPSM18} is a sound analyzer based on abstract interpretation that can reason about safety and robustness of DL systems. It trades precision for scalability and scales better than Reluplex. The precision of AI$^{2}$ depends on abstract domains used in the verification, and it might fail to prove a property when it actually holds. VERIVIS \cite{KYJS17} can verify safety properties of DL systems when attackers are constrained to modify the inputs only through given transformation functions. However, real-world transformations can be much more complex than the transformation functions considered in the paper.

\subsection{Attacks and Defenses of DL Systems}

A plethora of research has shown that deep learning systems can be fooled by applying carefully crafted adversarial perturbation added to the original input~\cite{goodfellow2014explaining,iclr18-a-boundary-attack,iclr18-ae-natural,iclr18-ae-spatial,szegedy2014intriguing,cw2017,carlini2017adversarial,chen2018automated}, many of which are based on gradient or optimization techniques.
However, it still lacks extensive study on how these adversarial techniques differentiate in terms of DNNs' internal states. In this study, we make an early attempt towards such a direction based on our proposed criteria.

With the rapid development of adversarial attack techniques, extensive studies have been performed to circumvent adversarial attacks. Galloway \etal~\cite{iclr18-ab-binarized} recently observe that low-precision DNNs exhibit improved robustness against some adversarial attacks. This is primarily due to the stochastic quantization in neural network weights.
Ensemble adversarial training~\cite{iclr18-ab-ensemble}, GAN based approaches~\cite{iclr18-b-defense-gan,iclr18-b-pixeldefend}, random resizing and random padding~\cite{iclr18-b-random}, game theory~\cite{iclr18-b-stochastic}, and differentiable certificate~\cite{iclr18-b-certified} methods are all investigated to defend against adversarial examples. 
By applying image transformations, such as total variance minimization and image quilting, very effective defenses can be achieved when the network is trained on the aforementioned transformed images \cite{iclr18-b-counter}. 
For more extensive discussion on current state-of-the-art defense techniques, we refer readers to~\cite{obfuscated-gradients}.

Our proposed testing criteria enable the quantitative measurement of different adversarial attack techniques from the software engineering perspective. This could be potentially helpful for understanding and interpreting DNNs' behaviors, based on which more effective DNN defense technique could be designed. In future work, it would be also interesting to examine  how to integrate the proposed testing criteria into the DL development life cycle towards building high quality DL systems.

\section{Conclusion and Future Work}

The wide adoption of DL systems, especially in many safety-critical areas, has posed a severe threat to its quality and generalization property. To effectively measure the testing adequacy and lay down the foundation to design effective DL testing techniques, we propose a set of testing criteria for DNNs. Our experiments on two well-known datasets, five DNNs with diverse complexity, and four state-of-the-art adversarial testing techniques
show that the tests generated by the adversarial techniques incur obvious increases of the coverage in terms of the metrics defined in the paper. This demonstrates that \emph{DeepGauge} could be a useful indicator for evaluating testing adequacy of DNNs.

To the best of our knowledge, our work is among the early studies to propose testing criteria for DL systems. We expect that the proposed testing criteria could be particularly amenable to DL testing in the wild. In the next step, we will continue to explore alternative testing criteria for DNNs, such as the combination of both hyperactive and hypoactive neurons. We also plan to study the proposed testing criteria guided automated test generation techniques for DNNs. We hope that our study not only provides an avenue to illuminate the nature and mechanism of DNNs, but also lays the foundation towards understanding and building generic and robust DL systems.

\balance
\bibliographystyle{ACM-Reference-Format}
\bibliography{ref}

\end{document}